\documentclass[trans]{IEEEtran}
\usepackage[T1]{fontenc}
\usepackage{hyperref}
\usepackage{url}
\usepackage{lettrine}
\usepackage{epsfig}
\usepackage{float}
\usepackage{footnote}
\usepackage{cite}
\usepackage{float}
\usepackage{soul}
\usepackage{amssymb}
\usepackage{amsthm}
\usepackage{amsmath,mathabx}
\usepackage{color}
\usepackage{epstopdf}
\usepackage{tabularx,tabulary}
\usepackage[font=footnotesize]{caption}
\usepackage{graphicx}
\usepackage{caption}
\usepackage{subcaption}
\usepackage{cuted}
\usepackage{dblfloatfix}  
\usepackage{bbm}
\usepackage{algorithm,algorithmic}
\usepackage{multirow}
\usepackage{wrapfig}

\usepackage[table,xcdraw]{xcolor}

\graphicspath{{../}}

\begin{document}
\title{When Wireless Communication Faces COVID-19: Combating the Pandemic and Saving the Economy}
\author{Nasir~Saeed,~\IEEEmembership{Senior Member,~IEEE,}  Ahmed Bader~\IEEEmembership{Senior Member,~IEEE,} Tareq~Y.~Al-Naffouri,~\IEEEmembership{Senior Member,~IEEE,}  and~Mohamed-Slim~Alouini,~\IEEEmembership{Fellow,~IEEE}
\thanks{Nasir Saeed,  Tareq~Y.~Al-Naffouri, and Mohamed-Slim~Alouini are with  King Abdullah University of Science and Technology (KAUST), Thuwal 23955-6900, Saudi Arabia (e-mail: nasir.saeed@kaust.edu.sa; tareq.alnaffouri@kaust.edu.sa; slim.alouini@kaust.edu.sa).  Ahmed Bader is with Insyab Wireless Limited, Dubai 1961, United Arab Emirates (e-mail: ahmed@insyab.com).}
}

\maketitle
\begin{abstract}
The year 2020 is experiencing a global health and economic crisis due to the COVID-19 pandemic. Countries across the world are using digital technologies to fight this global crisis, which, in one way or another, strongly relies on the availability of wireless communication systems.  Therefore, this paper aims to present the role of wireless communications in the COVID-19 pandemic from different perspectives. First, we show how wireless communication technologies are helping to combat this pandemic, including monitoring of the virus spread, enabling healthcare automation,  and allowing virtual education and conferencing. Also, we reveal the importance of digital inclusiveness in the pandemic and possible solutions to connect the unconnected. Next, we discuss the challenges faced by using wireless technologies, including privacy, security, and misinformation. Then, we present the importance of wireless technologies in the survival of the global economy, such as automation of industries and supply chain, e-commerce, and supporting occupations that are at risk. Finally, we reveal that how the technologies developed during the pandemic can be helpful in the post-pandemic era.
\end{abstract}

\section{Introduction}
History repeats itself, an infectious disease (COVID-19), once again attacking the human race. This time it is a respiratory virus that originated from the city of Wuhan, China, in December 2019 \cite{Na20}. COVID-19 affected almost all the countries in the World, significantly disrupting the noble Sustainable Development Goals (SDGs) of the United Nations (UN) \cite{Gallagher20}. Therefore the World Health Organization (WHO) declared it a global pandemic in March 2020. COVID-19 has made a lot of panic around the globe where the fights over the toilet papers look ugly enough; hopefully, we do not witness fighting for food during this pandemic. 

Since COVID-19 is a global issue, researchers around the world from different fields,  such as biomedical, virology, data analytics, and artificial intelligence, contributed to combating this pandemic.  In this context, more than 24 thousand research articles appeared online on COVID-19 in less than four months \cite{Hao20}. Moreover, in this global crisis, worldwide activities and businesses are heavily dependent on digital technologies {\cite{Ting20,  ITU20}. During this global emergency, applications of the digital technologies are numerous,  including tactile robotics to help the medical doctors and nurses at hospitals,  drones to monitor the crowds, artificial intelligence and deep learning for understanding the health-care trends, Internet of Things (IoT) for supply chain automation, and virtual learning to continue the education. All of these crucial applications and many more in this global pandemic rely on reliable and high-speed communication networks, putting tremendous pressure on these networks. In this context, the international telecommunication union (ITU) conducted an emergency meeting of the broadband commission for sustainable development that directed the governments, industries, and civil society to improve the capacity of communication networks at critical points, such as hospitals and transportation hubs  \cite{ITU_connectivity20}. Not only that, but they also emphasized on the importance of communication technologies in disseminating timely critical information, supporting e-learning for more than 1.5 billion students, training workers by digital means to improve productivity, and promoting e-businesses. 

With all this support of communication technology when we need it the most, it is also fighting with the on-going conspiracies of linking the fifth-generation (5G) networks with the spreading of COVID-19 \cite{Kaur20}. Based on this misinformation, people around the globe attacked and burnt the 5G towers \cite{guardian20}. To limit this false theory, researchers from the wireless communications community responded through various channels by showing the importance of 5G in the pandemic crisis and negating the health concerns over 5G operating frequencies  \cite{Ahmadi20, Popovski20, itu5g20}. 
 \begin{figure*}
\begin{center}
\includegraphics[width=1.8\columnwidth]{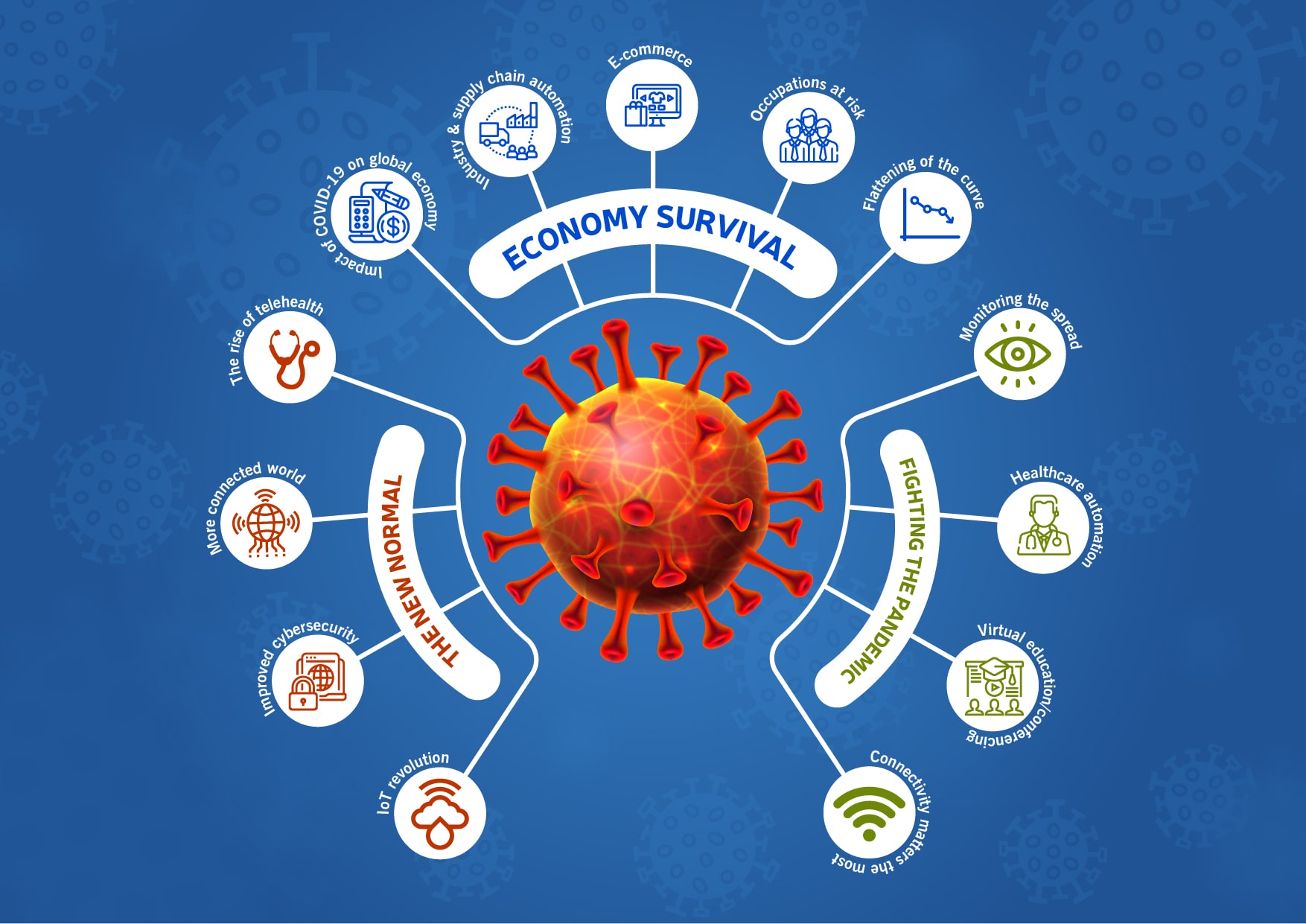}
\caption{Role of wireless communication technologies in the COVID-19 pandemic.}
\label{fig:wc}
\end{center}
\end{figure*}

Now that we agree on the significance of communication technologies in combating the COVID-19 pandemic let's look at ways that they are currently used and can further be improved. In this article, first, we show how the wireless communication technologies are assisting in fighting this pandemic, regarding monitoring of the crowd, contact tracing, supporting the medical doctors and nurses, enabling telemedicine, backing virtual learning, and providing resilient connectivity. We also show the importance of digital inclusiveness in the pandemic and possible solutions to connect the unconnected. Next, we discuss major challenges that emerge from the extensive use of wireless technologies, including privacy, security, and misinformation. Then, we uncover the decisive role of wireless communication technologies in saving the global economy downfall due to COVID-19, including automation of the industries and supply chain, assisting e-commerce, and supporting the occupations that are at most risk. Finally, we discuss extrapolating these technologies in the post-pandemic situation. In Fig. \ref{fig:wc}, we summarize the role of wireless communications in the COVID-19 pandemic.

\section{Fighting the Pandemic}
In the presence of this global pandemic, it is the top priority to prevent the spread of disease. Various wireless communication and positioning technologies have a significant impact on critical roles such as healthcare, education, and economy. In this section, we cover these technologies from various aspects, including  COVID-19 spread monitoring, healthcare automation, telemedicine, virtual learning, and digital inclusiveness, which are the biggest challenges at the moment in fighting this pandemic.

\subsection{Monitoring the Spread}
According to WHO, limiting mass gatherings is the best way to reduce the spread of COVID-19.  Besides mass gathering, social distancing is also crucial to minimize the spread of the virus. Both outdoor and indoor monitoring is essential to limit the transmission of the virus. In the following, we separately discuss various technologies that can achieve both outdoor and indoor tracking of the spread.

\paragraph*{\bf{Outdoor Monitoring}}
Various wireless communication and positioning technologies, such as drones, cellular positioning systems, and global positioning system (GPS), can be used to monitor the transmission of the virus in the outdoor environment. For instance, a network of drones can be of great benefit in monitoring crowds and maintain social distancing in metropolitan areas (see Fig. \ref{drone_monitor})  \cite{Richardson20}.  Various countries introduced "pandemic drones"  to enforce social distancing. Not only that, but these pandemic drones can also monitor temperature, flu, sneezing, and coughing in public places \cite{Kraut20, ians20, Meisenzahl20}. Once the drones collect the critical information, it needs to be communicated to the authorities in real-time for taking timely actions.  Pandemic drones with 5G connectivity can significantly help in such a situation due to its high transmission speed and low latency.
 \begin{figure}
\begin{center}
\includegraphics[width=0.8\columnwidth]{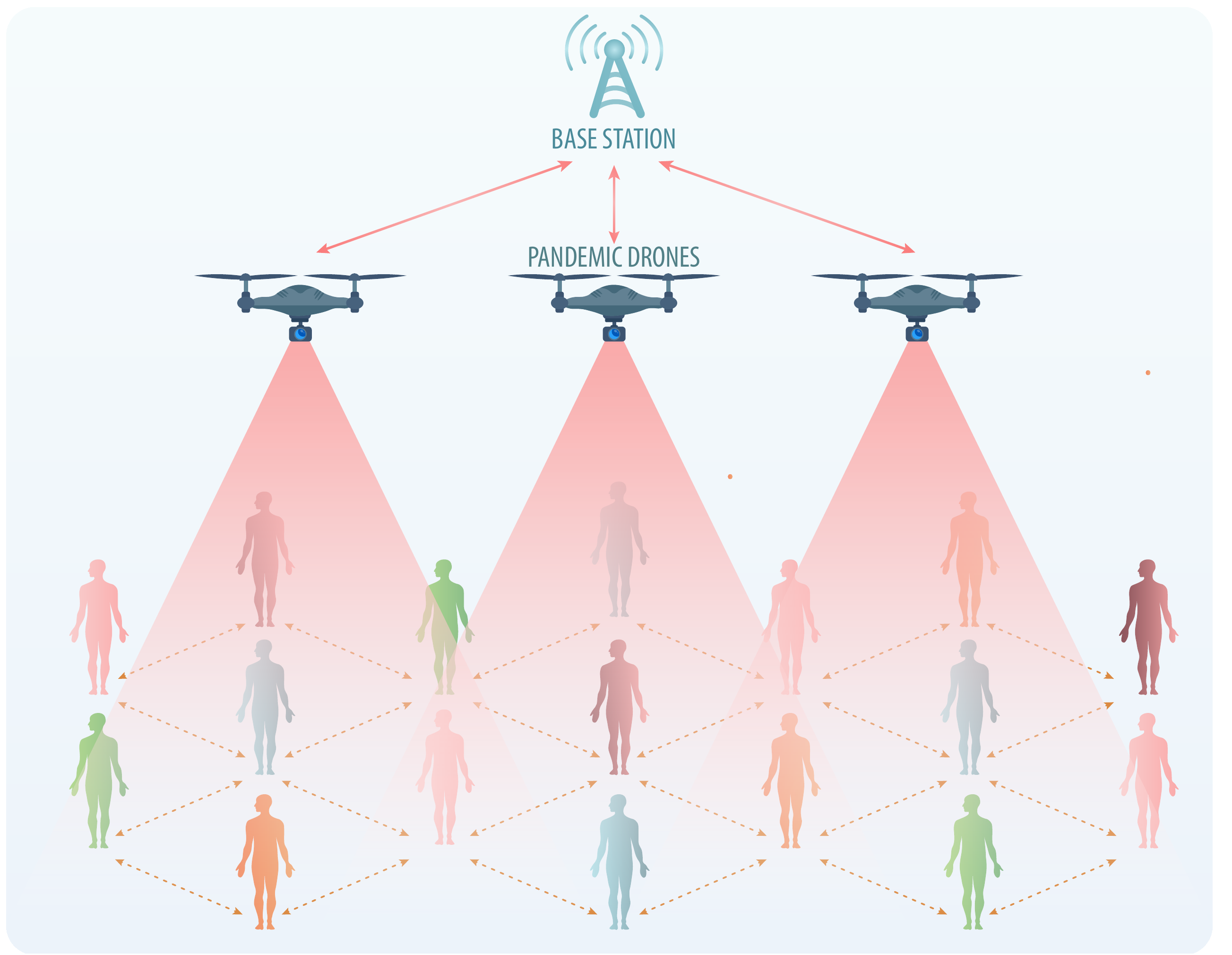}
 \caption{Connected drones monitoring social distancing.}\label{drone_monitor}
 \label{fig:com}
\end{center}
\end{figure}
Satellite communications is another major technology that is also of great benefit in monitoring, modeling, and responding to the COVID-19 spread.  The outbreaks of COVID-19 usually occur in a cluster in specific regions and then continue to spread. As such, the spread can be modeled using high-quality geospatial data. Also, the satellite imagery can identify the population that is at most risk and can track the health and testing facilities \cite{Zolli20}.

Another major issue in fighting this pandemic is contact tracing of the infected patients. In this context, Google and Facebook initiated a joint venture by sharing GPS-based mobility data of the users to the researchers \cite{Waltz20}. Also, Google and Apple are collaborating to build a contact tracing system that uses Bluetooth signals to sense the nearby smartphones and alert the users if they had encountered a possible COVID-19 patient.
Such geospatial data can be quite helpful for public health researchers to understand the spreading of the virus \cite{euhealth20}. For instance, the researchers from National Tsing Hua University used Facebook's data to show that limiting human mobility certainly reduces the transmission of the virus \cite{Chang2020}. With a slightly different approach, Kinsa Health distributed more than a million smart thermometers in the US to collect the temperature and geospatial information of the households remotely and predict the spread of the virus \cite{McNeil20}. The data from these intelligent thermometers can identify the clusters of patients with fever where a sudden spike can detect the presence of COVID-19. This way, an area of possible infected patients can be narrowed down, limiting the spread of the virus.  Analogously, researchers at John Hopkins University launched a mobile application where the user can add his/her body temperature and other symptoms that can be used by the government officials and health experts to take timely actions and predict future outbreaks \cite{Eisenberg20}.
 \begin{figure}[h]
\begin{center}
\includegraphics[width=0.95\columnwidth]{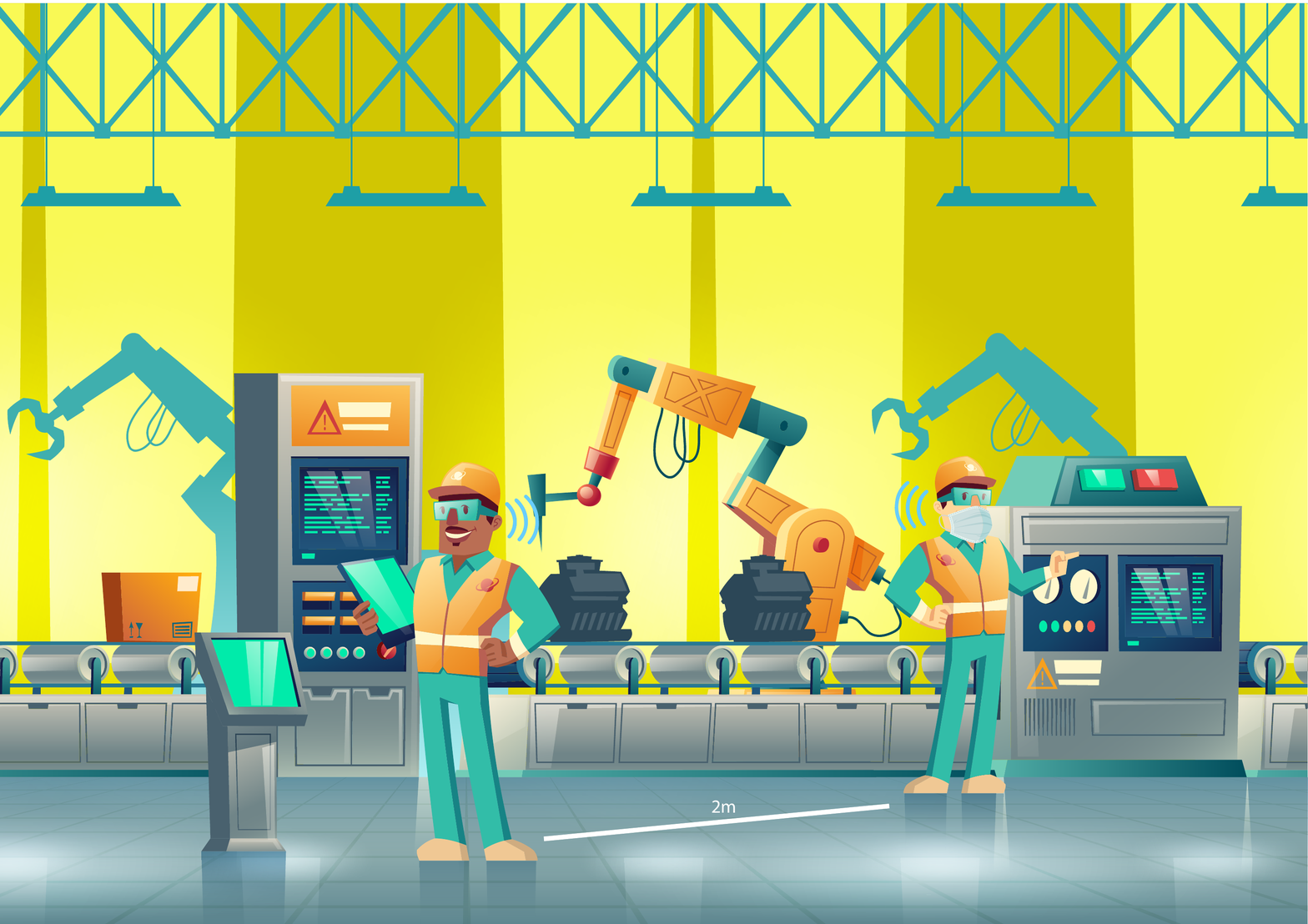}
 \caption{Proximity tracking for workers to maintain safe working distance.}\label{indoor_monitor}
\end{center}
\end{figure}
\paragraph*{\bf{Indoor Monitoring}}
 In the case of outdoor working environments, outdoor positioning technologies can maintain social distancing and safe working distance for the critical workforce.
However, in indoor environments, it may be more challenging due to the unavailability of GPS signals. Hence, novel solutions are needed to guarantee the monitoring of quarantined people and maintain safe working distance in indoor environments.  Various indoor positioning technologies such as radio frequency identification (RFID), Wi-Fi,  visible light, Zigbee, and Bluetooth can be promising solutions for monitoring the quarantined people and maintaining a safe working distance in indoor environments \cite{saeed2019mds}. One new solution recently developed for this cause is Proximity Trace, which is a sensing-and-communication technology for workforce distance monitoring while continuing essential businesses  \cite{PTrace20}. In Proximity Trace, a tag is attached to the worker hat or worn on a lanyard that transmits real-time alerts when the workers are in close contact (see Fig.~\ref{indoor_monitor}). There is a central gateway that collects all the workers' interaction information. This technology provides both passive and active solutions. In the passive approach, the interaction of a worker is tracked, in case, tested positive for the virus. In an active method, the workers are informed using sound or visuals to maintain the social distance.

Additionally, indoor localization and tracking methods that can take the sensing measurements from the users in proximity and create a network graph can be quite beneficial to maintain the safe working distance and perform contact tracing. For example, multidimensional scaling-based data visualization tools can be used to take proximity information as an input and create a network graph \cite{saeed2016robust, Khan2017, saeed2018}. These indoor monitoring solutions can help the workforce for industries eager to start regular operations while controlling the spread of COVID-19.

 \begin{figure}
\begin{center}
\includegraphics[width=0.95\columnwidth]{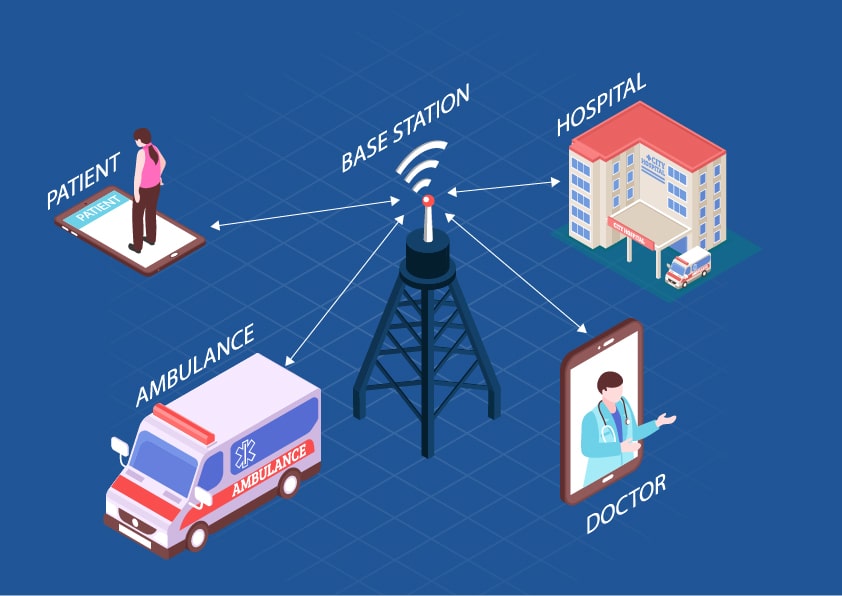}
 \caption{Concept of a wireless health monitoring system.}\label{telemedicine}
\end{center}
\end{figure}
\subsection{Healthcare Automation}
The outbreak of COVID-19  is proving to be a catalyst for 5G innovation, especially in healthcare automation. 5G-enabled medical robots can deliver drugs, check the temperature of patients, and disinfect rooms in the hospitals, reducing the exposure of medical personnel to the virus. These robots can cover the shortage of personal protective equipment required for doctors and nurses. For example, Ava Robotics' introduced iRobot for hospitals that help the doctors that are in quarantine at home to stay in the emergency rooms (virtually) without the risk of getting the virus \cite{Ackerman20}.
Not only this but also the hospitals can continuously collect the patient information using these connected robots and share it with the remote center that can significantly improve the efficiency of healthcare systems. The exchange of information from these robots and their navigation requires reliable and low latency communication that can be provided by 5G technologies. China has witnessed the use of 5G-enabled healthcare automation by implementing a 5G+ remote consultation system across its various hospitals \cite{Xiaoxia20}. In this system, the medical staff at a hospital can take consultation from the expert's miles away through remote video connections. Fig. \ref{telemedicine} illustrates an example of such a connected healthcare system.

\subsection{Virtual Education and Conferencing}
The COVID-19 pandemic is reshaping almost every aspect of our world, including ``Education''. Across the world, the schools have been closed for months, and it has forced the schools to take the classes online as social distancing measures continue. Online learning is the only option at the moment to continue education; however, online learning with cutting edge technologies can be a much better option compared to traditional offline classrooms. The essential requirements for e-learning include a good internet connection and access to a digital device; however, not all students and teachers have digital devices and high-speed connections, creating a digital divide problem. Moreover, a large portion of the global population has either limited or no internet access at all. To cope with this issue, governments around the globe need to distribute digital devices to low-income households and provide internet access \cite{Lageard20}. Regarding the connectivity issue, we discuss various solutions in section \ref{connectivity-issue}  that  can bring the internet to remote areas. 

Besides coverage of wireless networks, speed of the connection is of paramount importance in efficient e-learning; this is where 5G can play a significant role. 5G can even enable seamless augmented reality (AR) and virtual reality (VR) \cite{Ethirajulu20}, where students can take virtual classrooms (see Fig.~\ref{virtualreality}). 
\begin{figure}[h]
\begin{center}
\includegraphics[width=0.95\columnwidth]{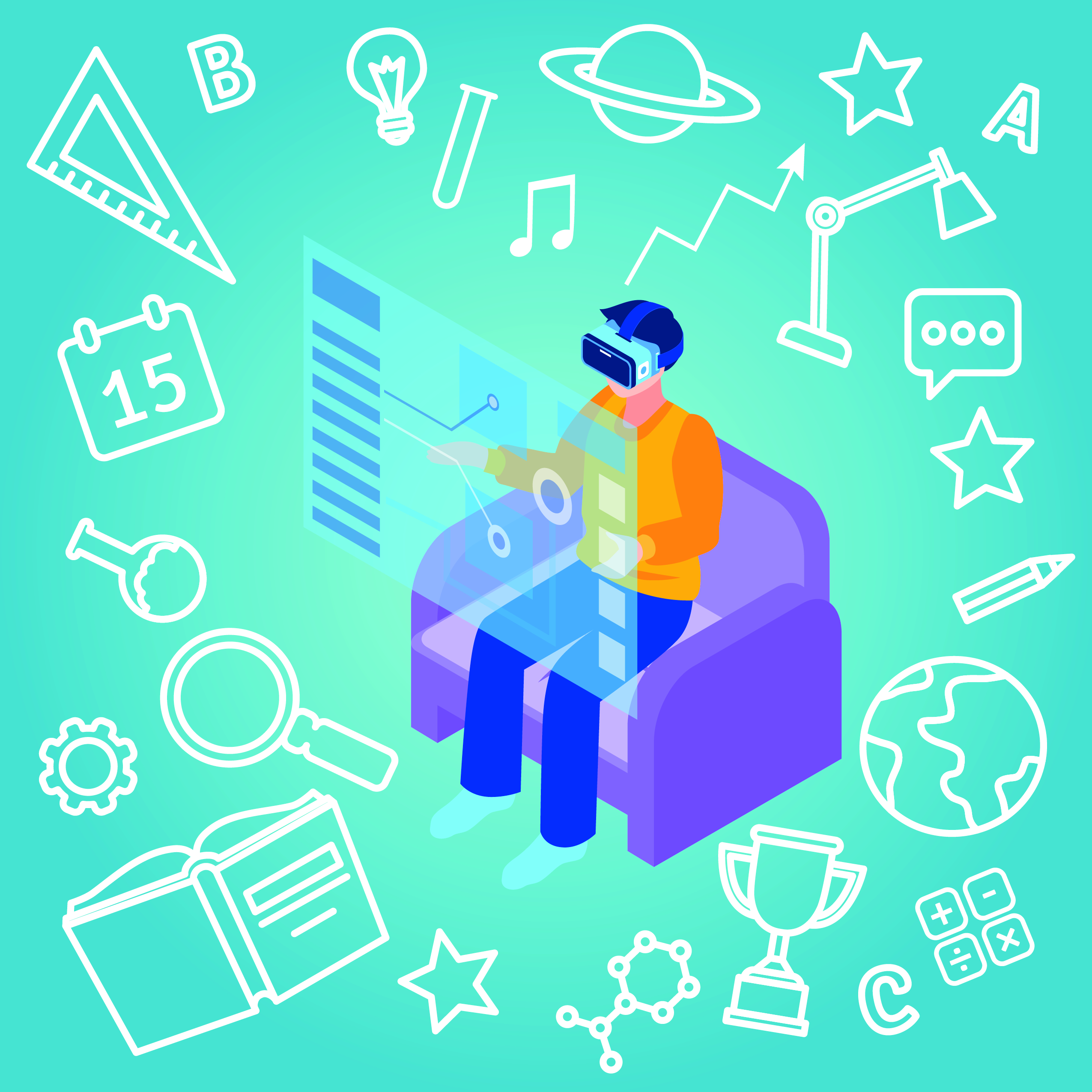}
 \caption{Concept of e-learning using AR/VR.}\label{virtualreality}
\end{center}
\end{figure}
In addition to virtual education, conferences and meeting across the world are becoming online. Again the videoconferencing requires a reliable and high-speed internet connection  \cite{Nordrum20}. Moreover, online conferences/meetings can be more enjoyable with the aid of VR. One such example is the recent conference of IEEE on virtual reality and 3D user interfaces \cite{Gent20}. Another recent example is the online virtual fabrication lab to help students training on the fabrication processes of semiconductor devices \cite{Hussain20}.  In summary, 5G technology can achieve these AR and VR solutions due to its extremely-low-latency, high bandwidth, and high reliability.


\subsection{Connectivity Matters the Most}\label{connectivity-issue}
The role of wireless communication technologies is not limited in containing the spread of virus and health automation; it is also helping people around the globe to be connected, making the social distancing a little less lonely. During the pandemic, connectivity matters more than ever because it enables the workforce to work remotely, students to continue their education, and promote e-business to deliver the basic needs \cite{Amon20}. This pandemic is validating the need and importance of 5G as the future of connectivity. 5G will enable telecommuting and virtual presence in a completely different way, integrate remote education as a part of everyday life, and allow remote healthcare and diagnostics \cite{ITU_rconnectivity20}. Moreover, the rapid change is user demands during this pandemic require the mobile network operators (MNOs) to switch from a ``competitive'' mode to a ``collaborative'' mode, providing the required peak capacity in certain areas. Also, the MNOs can collaborate closely with the governments to access any unallocated spectrum in the short-term to satisfy the increasing traffic demands.

Nevertheless, as mentioned earlier, it is the time to invest more in the digital inclusiveness solutions providing connectivity to the unconnected. This is one of the expectations from 6G technologies, providing low-cost connectivity solutions for rural and underprivileged areas \cite{dang2020, Yaacoub2020, saarnisaari20206g}. In the following, we briefly discuss these 6G solutions:

\paragraph*{\bf{Tethered Drones and Balloons}}
The new trend in wireless communications is enabling connectivity solutions from the sky. Tethered drones/balloons, also known as flying base stations, is one of these solutions that can provide connectivity in suburban and rural areas \cite{Reynolds20}. The drone/balloon is attached by a cable to ground, where the cable/tether maintains the drone/balloon position and also provides power to the on-board payload \cite{Kishk2020}. This way, the drone/balloon can stay for a longer time in the air, reducing the overall maintenance cost and providing uninterrupted connectivity \cite{kishk2019}. The drones/balloon-based solutions can either provide direct access to the user or work as a backhaul network for the terrestrial networks. 
 \begin{figure*}
\begin{center}
\includegraphics[width=2\columnwidth]{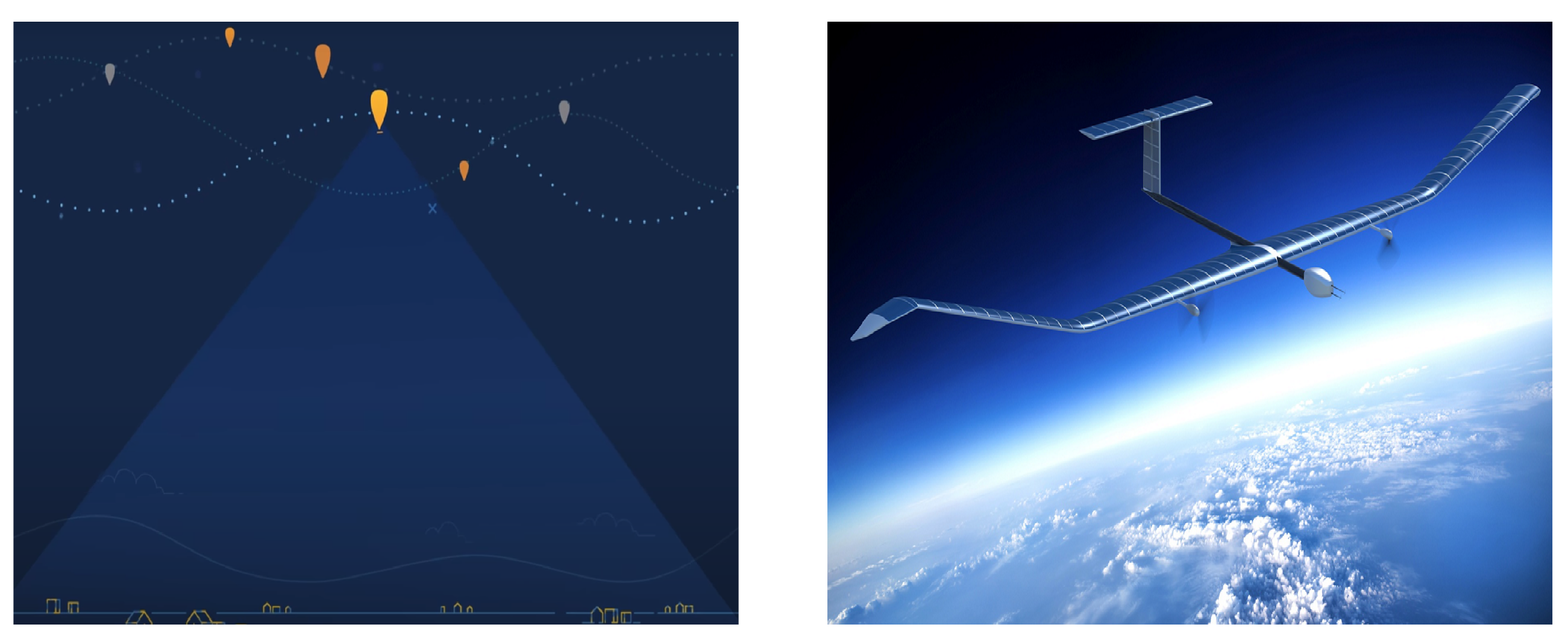}
 \caption{Balloons in Google's Loon project (on the left) \cite{Google20} and solar-powered drones in Zephyr project (on the right)  \cite{Zephyr20}, providing internet connectivity to underserved communities.}\label{connectivity}
\end{center}
\end{figure*}

\paragraph*{\bf{High Altitude Platforms (HAPs)}} The HAPs can provide connectivity to large rural areas due to its high altitude compared to the tethered drones/balloons. HAPs consists of balloons and drones positioned in the stratosphere at around 20 km. Recently, Google and Facebook invested in HAPs based solutions to bring internet to the remote areas of the world. One of the well-known projects on HAPs is Google's Loon (balloons), expanding the coverage of cellular networks for rural and underserved areas \cite{Google20}. Another big project was the Facebook Aquila project that consisted of solar-powered drones in the stratosphere, providing internet to the remote regions \cite{facebook20}. However, in 2018, this project was stopped by Facebook since it required aerospace industry expertise. Hence, Facebook collaborated with Airbus by using its solar-powered drone (Zephyr) to provide internet access to the rural areas \cite{Zephyr20}. Both these projects from Google and Facebook on HAPs can be promising solutions to connect the unconnected.

\paragraph*{\bf{Emerging Satellite Networks}}
In the space context, several industrial partners are joining the race to provide broadband internet access around the globe via thousands of satellites in low Earth orbit (LEO) \cite{Saeed2020}. Amazon’s Project Kuiper \cite{kuiper}, for example, is planning to deploy 3,236 LEO satellites at an altitude between 590 and 630 km, Starlink \cite{starlink} supported by SpaceX will install 12,000 LEO satellites, and OneWeb \cite{oneweb} and Telesat \cite{telesatleo} will place up to 900 satellites within the next few months. Such a large number of satellites has allowed mass production of components, thereby resulting in a significant reduction in satellite manufacturing costs. Finally, Astranis \cite{astranis} and Gapsat \cite{gapsat} are making efforts towards the massive deployment of mini-GEO satellites to improve satellite internet capacity worldwide.

Besides enabling or improving the connectivity in rural areas, there is a huge demand for broadband internet in urban and suburban regions. In general, buried underground fiber-optic cables provide broadband internet; however, it may not be a viable solution in a dense urban environment. Therefore, Terragraph, which is a Facebook technology, introduced an integrated backhaul access solution for the urban environment by taking the idea of routing in their fiber backbone and applying it into wireless networks \cite{terragraph}. Facebook is testing this technology worldwide, where it claims an average internet speed of 500 Mbps. One major challenge with the Terragraph technology is that it requires a strict line of sight conditions because it is operating at 60-GHz band.
\begin{table*}
\caption{Strategic response adaptation \cite{economy20}.}
\begin{tabular}{|p{4.0cm}|p{3.2cm}|p{3.2cm}|p{3.2cm}|p{2.5cm}|}
\hline
\textbf{Type of recovery response} & \textbf{\begin{tabular}[c]{@{}l@{}}Revenue growth\\ e.g., e-commerece\end{tabular}} & \textbf{\begin{tabular}[c]{@{}l@{}}Revenue loss (0-15 \%)\\ e.g., consumer goods\end{tabular}} & \textbf{\begin{tabular}[c]{@{}l@{}}Revenue loss (15-50 \%)\\ e.g., oil and gas\end{tabular}} & \textbf{\begin{tabular}[c]{@{}l@{}}Revenue loss\\  (+50 \%)\\ e.g., tourism\end{tabular}} \\ \hline
\textbf{V-curve (Recovery in few months)}                        & Boost in supply                                                                     & Push for fast back-to-normal                                                                   & Survive with slow back-to-normal                                                             & Set aside and look for a re-start                                                         \\ \hline
\textbf{U-curve (Recovery in 1-2 years)}                         & Nominal push for market share and growth                                            & Defend and look for new growth                                                                 & Look for innovation                                                                          & Look for innovation                                                                       \\ \hline
\textbf{L-curve (Fail to recover)}                               & Aggressive push for market share and growth                                         & Defend and look for new growth                                                                 & Re-invent or abandon                                                                         & Abandon                                                                                   \\ \hline
\end{tabular}
\end{table*}

\subsection{Challenges}
We cannot ignore the positive impact of wireless communication technologies in fighting the COVID-19 pandemic. However, it faces various challenges, such as privacy, security, and misinformation. In the following, we discuss each of these issues separately.

\paragraph*{\bf{Privacy}}
Although the contact tracking technologies can be quite helpful in curbing the spread of the virus, they indeed invade the privacy of the people \cite{McGee20}. The location information of users is accessed and monitored by these applications that can be used by governments as a surveillance tool. Worldwide human rights activists warn that the use of these applications in the pandemic may become a standard surveillance tool in the future. Governments need to answer a few questions without breaching the privacy of the users before implementing these applications. For example, Do the user know that his/her information is collected, and once the pandemic is over, the user can remove the data? How long will the governments retain the data collected for the COVID-19 pandemic?  Who will be able to access the user data and what are the rules to share the data. Analogously, the use of drones for monitoring social distancing or looking for sick people in public places will cause privacy issues. The aerial surveillance raises serious concerns regarding the right to privacy and freedom \cite{market20}.

 \paragraph*{\bf{Security}}
The unprecedented use of fixed and mobile broadband connections in the COVID-19 pandemic is causing a severe cybersecurity issue. According to the report of Akamai \cite{Leighton20}, the entire global internet traffic has increased by 30\% in March 2020, which is ten times more than average.
This record-breaking use of the internet creates vulnerability with the spike in malicious attacks, such as phishing emails, inserting malware to COVID-19 related resources, and even selling fake cures for COVID-19 \cite{Davies20}. Also, many enterprise employees work remotely from home, and it is challenging to provide authentication solutions for all users. As the businesses will become mainly online and follow e-commerce, more cyberattacks are expected \cite{Woollacott20}. Also, with the automation of industries where the devices are more connected than ever can initiate various vulnerabilities for cybercriminals. Perhaps, the new digital technologies that are evolving at a rapid rate during the pandemic can be classified as a high cybersecurity risk.

 \paragraph*{\bf{Is 5G Really Harmful? Combating the Misinformation}}
The COVID-19 pandemic has breathed new life into anti-5G conspiracy theories. Recently a British conspiracy theorist David Icke claimed in an interview, linking the COVID-19 outbreak and 5G technology. He argued that Wuhan was the first Chinese city to roll-out  5G, just before the virus outbreak. However, it is not true; Wuhan was among several Chinese cities, including Beijing, Shanghai, and Guangzhou, to get 5G. Also, the WHO found no evidence that 5G adversely affects health \cite{itu5g20}. The fear of 5G has prompted acts of violence in various countries where the cell towers have been attacked and set on fire \cite{guardian20}. These conspiracy theories have to be strongly invalidated (or adjust regulations) and need to be publicized in the media so that communication technology services keep progressing.
In this regard, the wireless communication technologies itself is combating with the misinformation. Researchers around the globe working with wireless communication technologies provided clarifications to fight this misinformation. For example, Peter Popovski, who is a professor of wireless communications at Aalborg University presented several points to clarify this misconception, such as fear of new technology, most of the wireless devices do not have 5G at the moment, and no evidence of  5G adversely affects on health  \cite{Popovski20}. To the best of our knowledge, also concluded by WHO, radiations of 5G and other communication technologies are not harmful to human health. Indeed, wireless communication technologies are vital in keeping the societies functioning when there is a massive lockdown due to COVID-19.

\section{Survival of the Economy}
The actions taken across the globe to control the spread of COVID-19 are extraordinary at the cost of unprecedented economic loss. There is great confusion going on among the policymakers about saving more lives or the economy. In this section, we cover the impact of COVID-19 on the global economy and show the role of wireless communication technologies in rescuing the economy.
\subsection{Impact of COVID-19 on the Global Economy}
Around the world, economies are crumbling due to the COVID-19 pandemic, where the price of every commodity has fallen significantly \cite{wbank20}.  The hard calculation for governments is trading life of the people who will die of the COVID-19 versus economic damage \cite{atkeson20}. Over the past three months, the world's biggest stock markets have turned red because the investors have to re-calculate the future for the economy. In the financial crisis (2007-08), there was a steady downfall of the economy. However, in the pandemic situation, investors are not pricing in due to strict lock-downs, leading to the crash of the global economy with low gross domestic product (GDP) growth this year (see Fig.~\ref{economy}). This all started when the Chinese government shut down the manufacturing industry to prevent the spread of the virus. Since China makes one-third of the World products, the shut down caused a delay in the production leading to the disruption in supply chain throughout the world \cite{mckibbin20}. For example, the car industry depends on various tools and machinery to manufacture their cars, where each one of these necessities experiences  1-3 months delay due to the shut down of the manufacturing sector in China. This means that car companies will not be able to launch new vehicles on time, missing out billions of dollars of sale. The same is true for other industries where the drop in sale can lead to the company's layoff and bankruptcies. One such example is the collapse of  OneWeb, a leading project on small satellites with 74 satellites already in orbits to provide internet from space \cite{Amos20}.
\begin{figure}[h]
\begin{center}
\includegraphics[width=0.9\columnwidth]{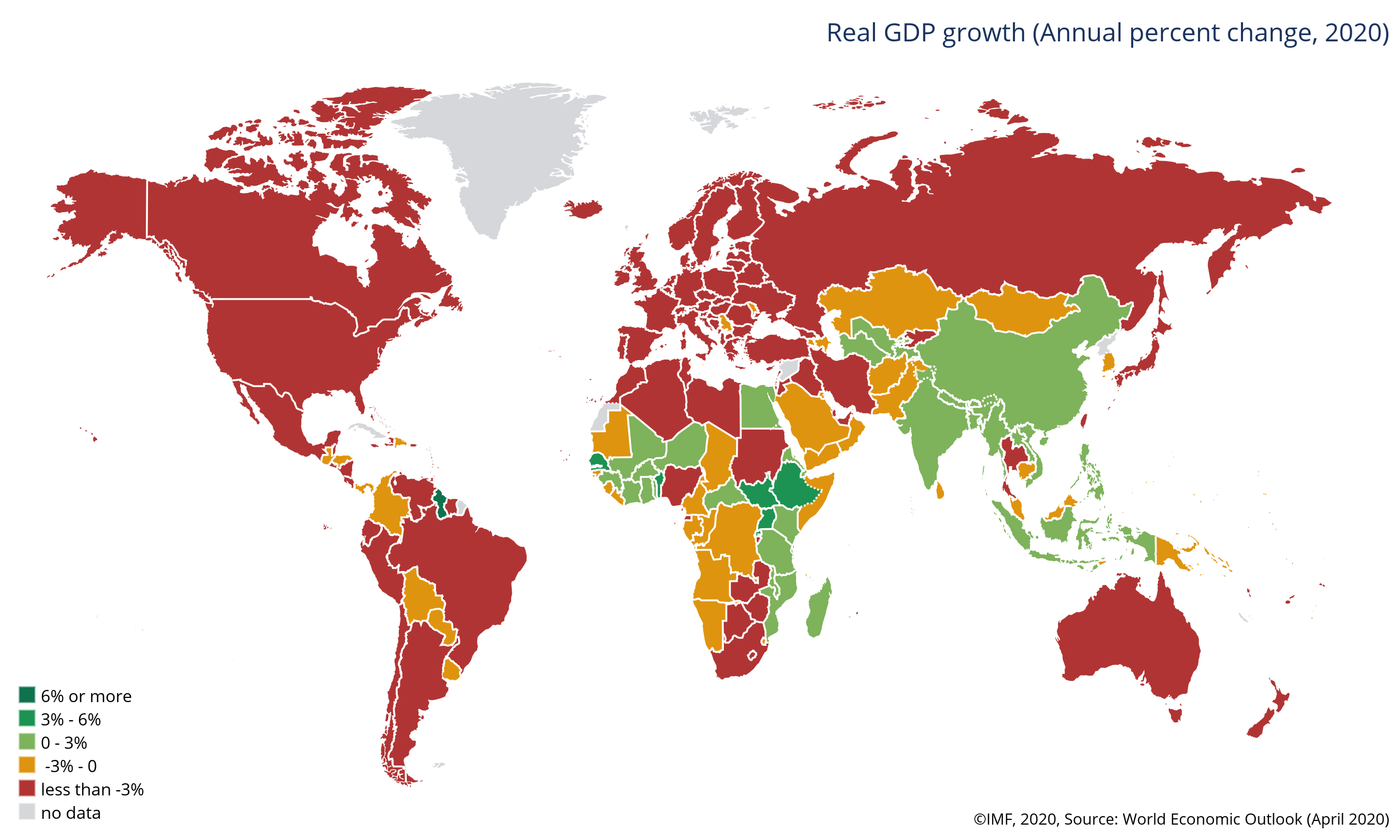}
 \caption{Global GDP downfall in 2020 due to the COVID-19 pandemic.}\label{economy}
\end{center}
\end{figure}
\begin{figure*}
\begin{center}
\includegraphics[width=2\columnwidth]{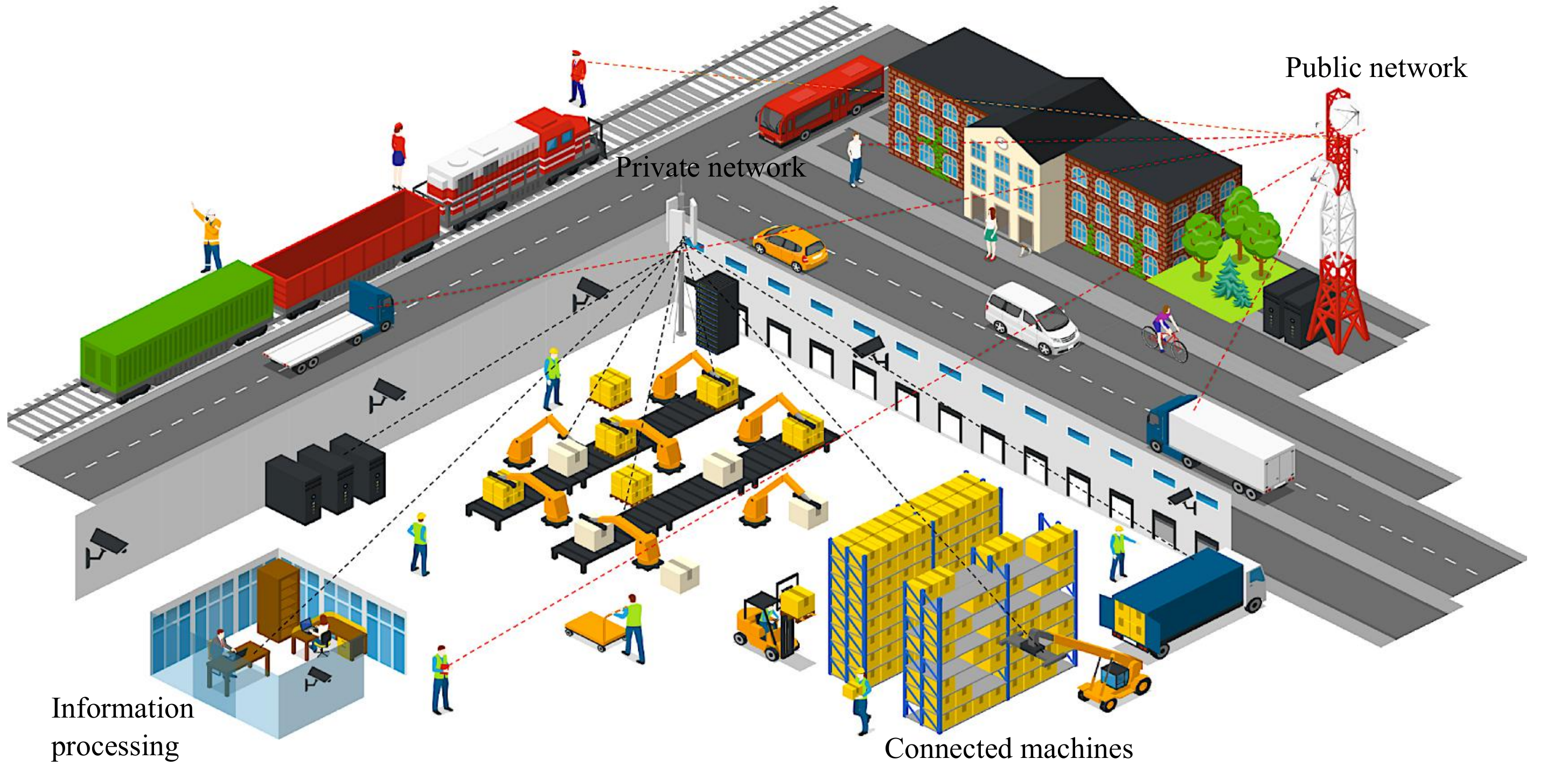}
 \caption{Illustration of an automated industrial network.}\label{iiot}
\end{center}
\end{figure*}

The industry needs to develop a strategic response by moving towards a low-touch or touch-less economy to fight this global issue and prepare for continuous disruptions. A matrix such as the COVID-19 strategy matrix can be followed to respond in such a crisis \cite{economy20}. For example, industries, such as e-commerce, can still generate good revenue in the short run and need only to keep up with the supply; however, sectors like tourism are very uncertain and need to look at a new start or abandoning in the long run. The COVID-19 strategy matrix is given in Table 1, which relates the impact and response for different industries. Moreover, it is the right time for the stakeholders to invest more in research and development (R\&D) because most of the innovative companies in the global financial crisis (2007-08) spent more on R\&D, which has paid off in profit and growth \cite{economy20}. Specifically, R\&D in digital technologies can pave the way for low-touch economy by developing novel smart solutions \cite{Field20}. Not only that, but these technologies can also monitor the state of the economy. For example, the researchers at WeBank have shown from the analysis of geospatial data that China's economy is recovering. They used visible and infrared images from the satellites to look for the activities in hot spots. Analogously, Ursa space systems is using images from their synthetic aperture radar satellites and use data analytics to see the impact of COVID-19 on global oil inventories \cite{Werner20}. With strict lockdowns, many companies are taking an interest in satellite imagery to collect information concerning operational activities in the concerned area.

\subsection{Automation of Industries and Supply Chain}
As mentioned earlier, one primary reason for the crumbling of global economies is the shutting down of industries and the break of the supply chain. However, recent advances in IoT can contribute to streamline the supply chain and reduce the impact of COVID-19 pandemic.  Also, IoT technology can play a significant role in the automation of industries that can further improve the supply chain. For instance, the Industrial Internet of Things (IIoT) can enable smart and adaptable manufacturing utilizing smart sensors and reliable connectivity for interconnecting and digitalizing of the traditional factories \cite{sisinni2018}. This can allow self-organizing, remote monitoring, and collaborative control for the conventional industries (see Fig. \ref{iiot}). The primary reason for establishing IIoT is that intelligent machines can efficiently collect, communicate, and act in a real-time industrial environment. To achieve this, IIoT requires stable and uninterrupted massive-and-ubiquitous high-speed connectivity, low-latency below one millisecond, and ultra-reliability to support real-time monitoring and provide automation.  Consequently, all these metrics are the characteristics of 5G technology, and therefore the concept of IIoT can be realized by using 5G \cite{cheng2018}. Moreover, 5G connectivity can potentially transform the supply chain by helping the companies to precisely measure consumer demands and react in real-time with changing situations.

 \begin{figure*}[t]
\begin{center}
\includegraphics[width=1.4\columnwidth]{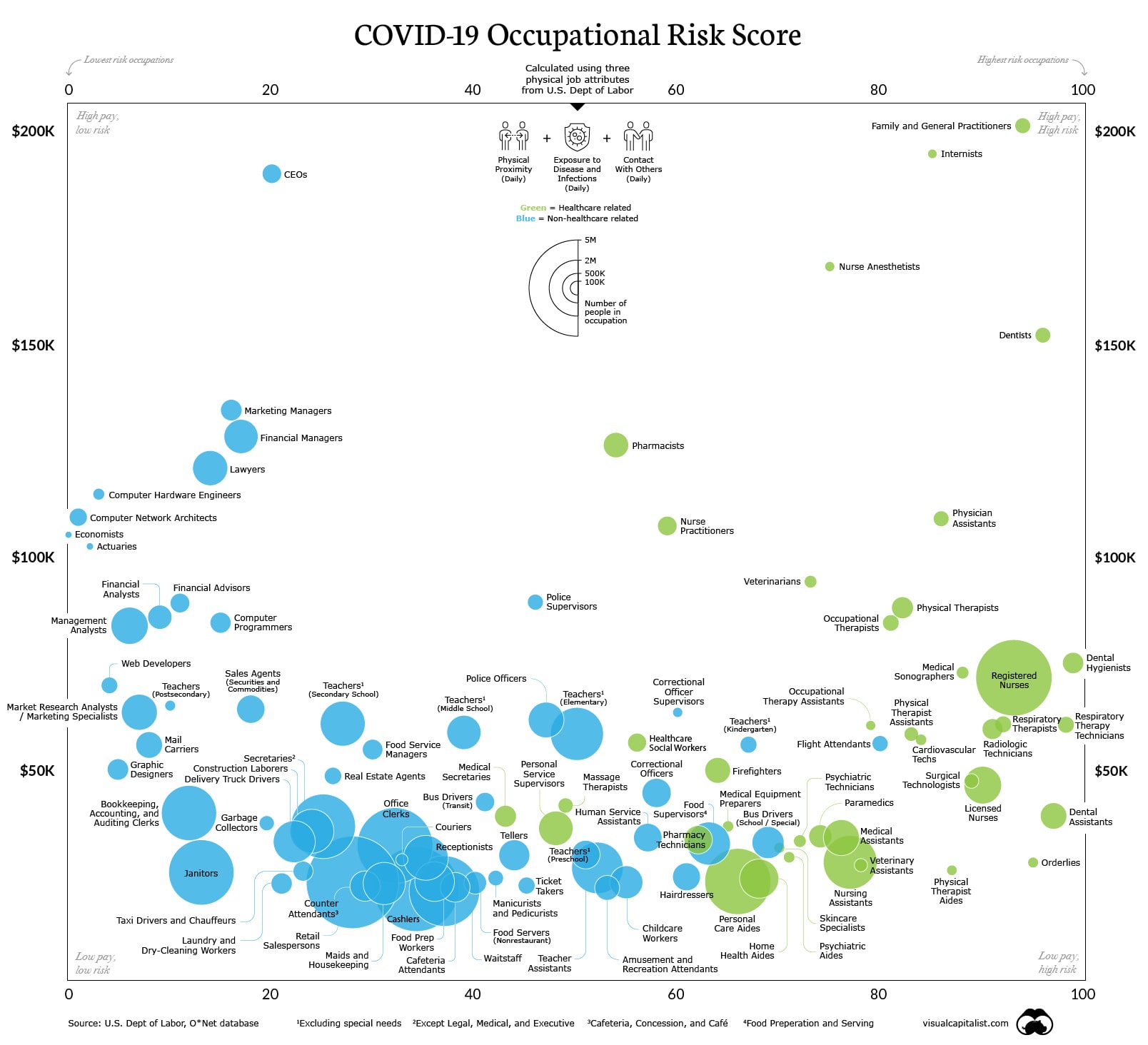}
 \caption{Occupations with respect to income and risk factor (0-100)  \cite{Lu20}.}\label{occupations}
 \label{fig:com}
\end{center}
\end{figure*}

\subsection{E-Commerce}
Due to the COVID-19 outbreak, e-commerce is emerging as a critical pillar in preserving jobs. For example, restaurants that are closed due to lockdowns can survive by starting online delivery services. Indeed, in some cases, e-commerce is creating new opportunities, e.g., Amazon recently announced to hire more  175,000 people to fulfill the customers' delivery demand and assist existing staff \cite{Amazon20}. Moreover, the combination of 5G and other technologies such as IoT,  AI, blockchain, and AR/VR will transform the e-commerce industry \cite{kshetri20185g}. For instance, with 5G connectivity,  IoT will enable the fast transfer of data to improve the consumer experience, track real-time inventory, and effectively manage orders. Likewise, AR/VR with 5G connectivity can allow the customers to place a real product virtually. With 5G, e-commerce can offer a unique high-resolution viewing experience for the enterprises, ultimately elevating their capacity.  A recent report of Adobe Digital claims that 5G will foster the revenue of e-commerce by \$12billion by the year 2021 \cite{Sterling20}, and it may further increase since most enterprises are moving towards e-commerce.
\subsection{Supporting the Occupations at Risk}
The current practice of remote working and keeping social distancing is luxury for some occupations, while people employed in sectors that provide essential services do not have this luxury. Hence, people who need to continue working as normal have a high risk of being exposed to the virus. Based on the attributes of the job, such as contacting others, close physical proximity, and exposure to the diseases, the US department of labor has defined an occupational risk score (see Fig. \ref{occupations}) \cite{Lu20}. According to this risk score, dental hygienists, bus drivers, and economists are at high, medium, and low risks, respectively. Digital technologies can play a significant role here in providing low-touch or touchless solutions for the occupations that are at medium/high risks. For example, the labor working in a supermarket can be equipped with smart wearable sensors to maintain minimum distances, avoiding the possible contamination of the virus. Similarly, telemedicine with the AR/VR capabilities can be used by dental hygienists for safe operations. This way, many occupations that are at risk can be protected, reducing unemployment.
\subsection{Flattening the Curve}
Across the globe, the terminology of “flatten the curve” is the focus of governments to reduce the spread of the virus, not taking into account the economic consequences.  However, every economic recession also plays a part in costing lives. For instance, unemployment is already rising; in the past few weeks in the US only, millions of people lost their jobs. Unemployment can further increase the death rate in a society giving rise to heart diseases. Besides unemployment, our modern societies require a sustainable health system, healthy nutrition, and clean water that may not be available during a deep recession. Hence, the long-term shutdowns can indeed flatten the curve but also would cause lives. Also, in low-income countries where the labor community survive on daily income, lockdowns can cause starving. According to the recent statement of the UN World Food Program (WFP), more prolonged shutdowns can lead to a hunger pandemic \cite{WFP20}. Therefore, epidemiological modeling of the COVID-19 spread and its socioeconomic impact is essential, where it follows an exponential growth in most countries around the world. Wireless communication technologies, as discussed in the previous sections, can undoubtedly help in flattening the curve that can further result in the opening of businesses.  Also, this will guide the policymakers to take actions regarding socioeconomic and health considerations.

\section{The New Normal}
The post-pandemic era or the new normal is going to be very different, bringing a new technological revolution. Although the digital technologies altogether could not end the pandemic yet, it is diving deeper to discover its true potential in different sectors. In this section, we cover such wireless communication technologies that are developed during this pandemic and can be used in the post-pandemic era.

\subsection{The Rise of Telehealth}
COVID-19 is the game-changer for the healthcare industry. According to Statista, the forecasted revenue of the global telehealth sector is around \$332.7 billion by 2025 \cite{Statista20}, where it cannot be achieved without efficient and reliable wireless connectivity. The role of IoT, robotics, and AR/VR with high-speed connectivity will truly revolutionalize the healthcare industry. It will not only enable the medical specialists to analyze the patients' data but also contribute to remote operations \cite{Bernard20}. Moreover, it will also reduce the travel expenses since the best doctors from anywhere in the world can be accessed online. The telehealth can also keep those clinicians productive who are unable to offer the patients an in-person visit.

\subsection{More Connected World}
The COVID-19 pandemic is showing us the true potential of connectivity for people and interconnected technologies. Across the globe, the use of audio/video applications, such as Google Hangout, Zoom, and Skype, are keeping the people connected and allows e-learning/teleconferencing. The dependency on these applications in the COVID-19 can lead to more remote working and a blended education system in the future. Moreover, the blend of 5G and AR/VR can lead to virtual labs, classrooms, and conferencing.

Similarly, the solutions developed to provide connectivity to the rural areas during this pandemic can lead to overcoming the digital divide problem in the long run.  
 
 \subsection{Improved Cybersecurity}
Besides threatening the healthcare systems and economy, the COVID-19 pandemic is causing serious cybersecurity issues. This is mainly due to more dependency on the internet in daily life while working remotely in global lockdowns. This dependency creates vulnerability with the spike in malicious attacks, such as phishing emails, inserting malware to COVID-19 related resources, and even selling fake cures for COVID-19 \cite{Davies20}. Moreover, as the businesses will become mainly online and follow e-commerce, more cyberattacks are expected \cite{Woollacott20}. Also, with the automation of industries where the devices are more connected than ever can initiate various vulnerabilities for cybercriminals. Amidst "flatten the curve" for the spread of the virus, increase in the cybersecurity threats should also be mitigated. With this in mind, the awareness and tools developed during the pandemic for cybersecurity can help online businesses become more robust in the new normal.

\subsection{The IoT Revolution}
The IoT networks are seeing astonishing growth in the COVID-19 era, which will result in the automation of many sectors, including health, manufacturing, consumer electronics, and transportation.  The role of IoT will further intensify with the rollout of 5G, leading to the applications beyond our imagination from smart cities and smart oceans to smart space \cite{saeed2019towards}. This fast-growing research in the IoT field will also help the human race to be more prepared for natural calamities and pandemics in the future.
 
 \section{Conclusion}
Wireless communication technologies are playing a significant role in winning the fight against the age of disruption and looking forward to the new normal. In this paper, we discussed different global challenges that originated due to the COVID-19 pandemic, and then we uncovered the fact that wireless communication technologies are necessary to fight these challenges, such as monitoring the spread of the virus, enabling healthcare automation,  and allowing virtual education and conferencing. Moreover, we have also discussed the challenges faced by these technologies, such as privacy, security, and misinformation. Additionally, we also reveal that wireless communication technologies are assisting in the survival of the global economy by helping in different sectors, such as industries, supply chains, and e-commerce. Finally, we highlight a few of the use cases that will revolutionize during this pandemic, leading to a technological transformation in the future.


\bibliographystyle{IEEEtran}
\bibliography{nasir_ref}

%
%
%

%

\end{document}